\documentclass[12pt,preprint]{aastex}

\newcommand{\be}{\begin{equation}}
\newcommand{\ee}{\end{equation}}
\newcommand{\bea}{\begin{eqnarray}}
\newcommand{\eea}{\end{eqnarray}}
\def\ceqn(#1){equation~(\ref{#1})}
\def\ceq(#1){(\ref{#1})}
\def\eqn(#1){\label{#1}}

\begin{document}
\title{Search for high-frequency periodicities 
in time-tagged event data from gamma ray bursts
and soft gamma repeaters}

\author{Adam T. Kruger,\altaffilmark{1}  Thomas J. Loredo,
\altaffilmark{2} and Ira Wasserman\altaffilmark{2,3}}
\altaffiltext{1}{Department of Physics,
Cornell University, Ithaca, NY 14853}
\altaffiltext{2}{Center for Radiophysics and Space Research,
Cornell University, Ithaca, NY 14853-6801}
\altaffiltext{3}{Floyd R.\ Newman Laboratory of Nuclear Studies,
Cornell University, Ithaca, NY 14853-5001}

\begin{abstract}

We analyze the Time-Tagged Event (TTE) data from observations of gamma
ray bursts (GRBs) and soft gamma repeaters (SGRs) by the Burst and
Transient Source Experiment (BATSE).  These data provide the best
available time resolution for GRBs and SGRs.  We have performed an
extensive search for weak periodic signals in the frequency range 400
Hz to 2500 Hz using the burst records for 2203 GRBs and 152 SGR
flares.  The study employs the Rayleigh power as a test statistic to
evaluate the evidence for periodic emissions. We find no evidence of
periodic emissions from these events at these frequencies.  In all but
a very few cases the maximum power values obtained are consistent with
what would be expected by chance from a non-periodic signal. In those
few instances where there is marginal evidence for periodicity there
are problems with the data that  cast doubt on the reality of the
signal. For classical GRBs, the largest Rayleigh power occurs in bursts
whose TTE data appear to be corrupted. For SGRs, our largest Rayleigh
power, with a significance of $\approx 1$\%, occurs in one record for
SGR~1900$+$14 (at $\approx 2497$~Hz), and in no other outbursts
associated with this source; we thus consider it unlikely to represent
detection of a real periodicity.  From simulations, we deduce that the
Rayleigh test would have detected significant oscillations with
relative amplitude $\approx 10$\% about half the time. Thus, we
conclude that high frequency oscillations, if present, must have small
relative amplitudes.

\end{abstract}

\keywords{gamma rays: bursts --- methods: statistical}

\section{Motivation}
\label{sec:motivation}

In the last decade observations of classical gamma ray bursts
(GRBs) and soft gamma repeaters (SGRs) have finally identified the 
sites of the sources of these high energy transients, yet the
physical nature of the sources remains mysterious 
(Paczy\'nski 1995;
Costa et al. 1997; Bond 1997; Djorgovski et al. 1997; 
M\'esz\'aros \& Rees 1997; Horth et al. 1998; 
Frail et al. 1998; Taylor et al. 1998; Kulkarni et al. 1999).
Detailed
timing analysis of the high energy emission from these sources
offers important clues regarding the nature of the central engines. 

The positive detection of periodic emission in SGR flares suggests a
rotating neutron star origin for these events, but to date, only two
events have shown evidence of periodic emission (Barat et al.\ 1979;
Cline et al.\ 1980; Terrell et al.\ 1980; Barat et al.\ 1983;
Kouveliotou et al.\ 1999).  The current picture of SGRs was first
introduced in 1992 when it was proposed that they are neutron stars
with extremely large magnetic fields, coined ``magnetars'' (Duncan and
Thompson 1992). Part of the motivation for this model is the 8 s
periodicity observed in the 5 March 1979 event, thought to be the
rotational period of a neutron star associated with the source.  In
addition to rotation and precession, faster nonradial pulsations may be
involved in the SGR events (McDermott, Van Horn and Hansen 1988; Duncan
1998).  To date, no periodic emission has been detected from
(classical) GRBs, but it might be expected at some level in some
models.  If binary compact stars or matter orbiting near neutron stars
or black holes are involved in GRBs, it is likely that there are rapid,
possibly periodic, associated phenomena.  Quasi-periodic oscillations
(QPOs) have been observed from accreting neutron star systems with
frequencies as high as 1200 Hz. Some QPOs are thought to be associated
with gas orbiting a neutron star at the innermost stable circular orbit
at $r= 6GM/c^2$, where the Kepler frequency is $2200 (M_\odot/M)$~Hz
for a star of mass $M$; for a black hole it may be as high as $11300
(M_\odot/M)$~Hz.  Recently, significant variability on time scales
even shorter than 1~ms has been seen from Cen X-3, attributed to photon
bubble oscillations that could produce quasiperiodic variability (Klein
et al. 1996; Jernigan, Klein \& Arons 2000).  While it is presently
unclear whether there are multiple classes of GRB progenitors, most
plausible models involve a central black hole and a (temporary) debris
torus around it. This includes massive progenitor systems, such as
hypernovae or collapsars where a massive rotating star collapses to a
black hole leaving behind an accreting disk of material that releases
energy (Paczy\'nski 1999; Nomoto et al. 2000), and many models
involving the merger of compact stars. Short time scale variability,
possibly including high frequency ($f\approx 10^{3}$ Hz) pulsations is
expected for these types of sources, where material may be orbiting
close to a nascent neutron star or black hole.

Further observation of periodic emission in SGR flares 
or detection of periodic emission from GRBs
could provide 
valuable clues about the nature of the engines powering these events.
High-resolution, 
time-tagged data which makes a search for high frequency
periodicities possible 
is available through the BATSE data archive
at the Compton Observatory Science Support Center (COSSC).\footnote[1]{%
http://cossc.gsfc.nasa.gov/}
We know of no other
large-scale search for periodic emission that examines these data. 

\section{TTE data type }
\label{sec:data}

BATSE observed thousands of gamma ray bursts during the nine year 
life of the {\it Compton Gamma Ray Observatory} ({\it CGRO}).  
In this study we 
analyze the BATSE Time-Tagged Event (TTE) data for 2203 GRBs and 152 
outbursts from SGRs. The BATSE TTE data type consists of arrival 
times (with 2 $\mu$s resolution) for $2^{15}$ individual Large Area 
Detector (LAD) photon detection events for each detected burst. 
BATSE will generate a burst trigger if the count rate in two or 
more detectors exceeds a threshold specified in units of standard 
deviations above background (nominally 5.5). The rates are tested 
on three time scales: 64 ms, 256 ms and 1024 ms.  TTE data is
stored in a ring buffer that constantly records data from
all eight LADs until a trigger.  Once a trigger occurs, the
most recent $1/4$ of the buffer is kept, and the remaining
$3/4$ of the buffer is filled with events from the triggered
detectors (or for the four highest rate detectors if more
than four were triggered).  In our analysis, we use data only
from triggered detectors (i.e., we ignore some of the pre-trigger data).
The TTE data type has the best time resolution of any BATSE burst data
and thus provides the most precise information available about burst
temporal behavior. Since burst intensities vary and TTE data is
collected from a subset of the LAD detectors after trigger, the TTE
data sets vary in time coverage and in number of arrival times from
triggered detectors. In this study the number of arrival times analyzed
from a single record varied significantly, with a maximum number of
29354.  Under typical (non-burst) conditions the background gamma-ray
photons are detected at a rate which fills the available buffer in
about 2~s. In spite of this expectation, there are a number of TTE
records which cover $\approx 3.75$~s, and a few which last an
anomalously long time ($\approx 7-8$~s). In the analysis we account for
these variations to determine the significance of candidate
periodicities.

In addition to these characteristics of the data sets, there is also
significant variation in the burst profiles themselves evident from
visual inspection of binned light curves.  Some of the qualitative
features of the light curves are useful for categorizing the data, such
as the number of peaks and the fraction of the burst light curve covered by
the TTE data.  Since the buffer for TTE data is limited in size, many
bursts are not fully covered. Since 1995 it has been recognized that
the distribution of T90 values for GRBs---the duration spanned by the
central 90\% of the burst photons---indicates two classes of bursts:
one population with durations between 0.1 s to 2.0 s and another from 2
s to 100 s (Fishman \& Meegan 1995).  For our analysis we do not treat
these populations separately. We do treat SGRs and GRBs
separately, but we do not distinguish between long and short bursts. 

The COSSC maintains an easily accessible collection of TTE data for short bursts.%
\footnote{http://cossc.gsfc.nasa.gov/batse/batseburst/tte/}
The collection is restricted to those bursts for which the TTE record
covers the entire burst duration.  It is possible that periodic
emission occurs during burst precursors or the during the onset of
bursts.  For this reason we do not limit ourselves to TTE records that
fully cover the burst; we obtain TTE data from the full COSSC FTP
archive of BATSE data in FITS format.   Also, for a small number of
bursts, inspection of the TTE data by the COSSC team indicates that the
triggered detectors were misidentified by the TTE hardware.  For these
triggers we have used only events from the true triggered detectors,
which are sometimes limited in number.  We note below some other rare
anomalies among the TTE data.

\section{The Rayleigh Test}
\label{sec:Rayleigh}

The most widely-used procedure for determining which modes are 
present in a process such as photon emission is analysis of
the power spectrum calculated from the discrete Fourier transform (DFT)
of uniformly sampled data. Application of the DFT to arrival
time series data requires binning of the data to produce
equally spaced samples.  
Binning is a subjective procedure; choice of bin width can affect
the apparent significance of a detection and limits sensitivity
on short time scales.
To avoid subjectivity and loss of sensitivity associated with
binning, we use a procedure which does not require binning.
We conduct 
our search for periodicities by using the Rayleigh power, $R$, as a 
test statistic (Mardia 1972; Lewis 1994; Mardia and Jupp 2000). 
The Rayleigh test was developed to detect a preferred direction in
circular data (angles spanning [0, 2$\pi$]).  For data such as the
BATSE TTE data, one can assign to each datum a phase $\theta_{i}(f)$
given by
\begin{equation}
\theta_{i}(f)=[2 \pi f t_{i}] \mbox{mod} 2 \pi,
\label{theta}
\end{equation}
where the ${t_{i}}$ are the photon arrival times and $f$ is a 
candidate frequency. The Rayleigh power for a trial frequency $f$ 
is calculated from the set of $m$ photon arrival times \{$t_i$\} by
\begin{equation}
R(f) = \frac{1}{m}\left[\sum_{i = 1}^m \sin(2\pi f t_i)\right]^2 + \frac{1}{m}\left[\sum_{i = 1}^m \cos(2 \pi f t_i)\right]^2.
\end{equation}
(Note that there are two normalization conventions for $R$ in the
literature, the one we use, and one that divides our $R$ by 2.) If one
takes each event as defining a unit vector in the plane with components
($\cos \theta_{i}, \sin \theta_{i}$), then $mR$ is the magnitude of the
sum of all the vectors.  If there is a preferred phase present in the
data, the sum will line up in the corresponding direction. If a
particular set of measurements yield a resultant vector with a large
magnitude for a particular choice of frequency, a periodic signal may
be presetn at that frequency.

To quantify the evidence for a periodicity the Rayleigh test uses the
Rayleigh power to test the null hypothesis that there is no preferred
phase; i.e., that the distribution of $\theta_{i}$ is uniform. 
Under this hypothesis
asymptotically $2R$ is the sum of the squares of two 
independent standard normal random variables with unit standard 
deviation and it follows that the random variable $R$ is asymptotically 
distributed as $\chi^2_2$, so that the probability that $R$ exceeds 
some threshold $R_0$ is
\begin{equation}
\mbox{Prob}(R > R_0) = e^{-R_0}.
\label{chisqr}
\end{equation}                             
If a single candidate frequency is specified a priori, equation 
(\ref{chisqr}) can be used to assess the evidence for a periodicity
at that frequency. If the frequency is unknown, one must evaluate
$R(f)$ at many trial frequencies to find the best candidate frequency
(the one with the largest $R(f)$). The significance associated with
the candidate frequency must account for the fact  that many 
(possibly dependent) $R(f)$ values were examined. Finally, if one 
examines many data sets for evidence, the probability of finding
a significant departure from the the predictions of the null
simply due to chance increases. This must be taken into account
when searching a catalog of events. We now discuss how we established
a search strategy and how we accounted for the strategy in
calculating significances.


\section{Frequency search range and oversampling}

In this section we will discuss how the 400 Hz lower bound for our
frequency search was chosen and how the number and spacing of
the frequencies searched affects the distribution of values of
$R(f)$ expected from a burst record.  For clarity we will refer to a
particular event, BATSE trigger  6659, GRB 980326, which has been
considered to be a hypernova candidate (Nomoto 2000). The light curve
for this burst is shown in the inset to Figure 1.  

\begin{figure}
\plotone{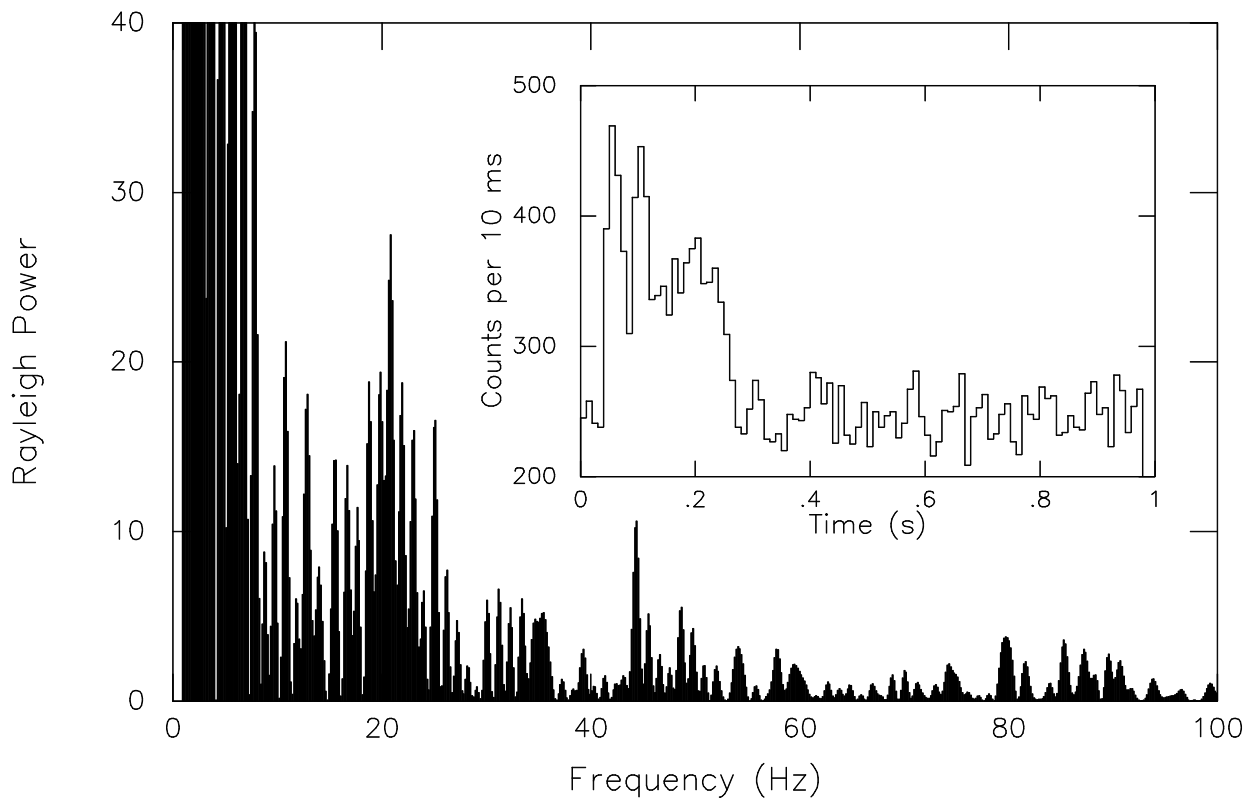}
\caption{Power spectrum for BATSE trigger 6659 with
frequencies sampled at rate 2$\pi$ greater than the independent 
frequency sampling rate. This figure 
demonstrates how large calculated Rayleigh power can appear at low
frequency due to 
long-timescale features of the burst. The light curve for this burst
is shown in the inset with the photon arrivals binned in 10 ms bins. 
The onset of the burst is characterized by two brief 
spikes which result in a large power value near 20 Hz.}
\end{figure}

For the results of
the Rayleigh test to be meaningful, the null hypothesis must be a
reasonable description of the data in the absence of a periodicity.
One way to have uniform phases in absence of a periodicity is with a
flat light curve, but this is not reasonable in this case.  In our analysis,
the uniform phase hypothesis is made reasonable through our choice of
frequency search range.

We note that the presence of the burst envelope can lead to large power
values at frequencies corresponding to time scales in the envelope.
Power due to major features of the burst can give an initial indication
that there is some periodic component to the emission.  This is
apparent in the Rayleigh power spectrum shown in Figure~1.  It is clear
the large power at 20 Hz arises due to the two large spikes observed in
the light curve at roughly 0.05 sec and 0.1 sec. This very large
Rayleigh power does indicate that the photons do not come in uniformly
during the observation time, but two spikes do not constitute a
periodicity.  We wish to avoid the influence of the burst envelope in
our study.  Figure 1 also shows that power due to the major
burst features is evident out to frequencies of nearly 60 Hz, but at
frequencies higher than this, the values are typical of background.
Many of the bursts in our study have multiple narrow spikes, and for
some of them power from the envelope extends to even higher
frequencies. By examining the Rayleigh power for bursts in our sample
at low frequencies, as well as for simulations of bursts with a
smoothly varying envelope function modeled piecewise as a two-sided
exponential, we set a conservative lower limit for our frequency search
of 400 Hz. By choosing this lower limit we eliminate the power due to
long-time structure in the burst envelope, and can use the Rayleigh
test to search for high-frequency periodic emission with confidence.
As a high-frequency cutoff for our search we chose 2500~Hz, a value
just above the rotational break-up frequency for a $\sim 10$~km,
1.4~$M_{\odot}$ neutron star.

In Figure 1 we have sampled the frequency axis at a rate $2\pi$ higher
than if we limited ourselves to frequencies such that an integer number
of periods is spanned by the data (Fourier frequencies).  This provides
us with a greater opportunity to observe large Rayleigh values if they
are present since the peaks of $R(f)$ can lie between Fourier
frequencies.  The spike in the power spectrum near 20 Hz reaches 27.5,
whereas calculations of power at only Fourier frequencies results in a
maximum of 15.0.  Under the null hypothesis, the significance level
associated with a power of 27.5 is formally over $2\times 10^{5}$ times
smaller than that associated with 15.0. But more frequencies had to be
examined to find the larger value, and this must be accounted for to
assess its true significance.

If one 
examines only Fourier frequencies, the resultant powers will be 
statistically
independent, but spacing of the samples may lead one 
to miss large values of the Rayleigh power. 
The benefits of higher
resolution outweigh the significance penalty for oversampling.
We chose to oversample using the same 
number of frequencies for each
record. We compensate for this oversampling by 
determining an effective number of independent frequencies
sampled using simulations. We have done this by finding the 
maximum value of the Rayleigh power achieved in our calculations 
of the power at 13194 frequencies between 400 Hz and 2500 Hz. 
Restricting ourselves to independent samples would lead to 
$2100 T$
test frequencies, where $T$ is the time coverage of the record, which
is typically $1-3$ s.  We are thus oversampling by factors typically
ranging from two to six (similar to recommendations in the literature,
e.g., Lewis 1994).  The probability distribution for the maximum power
expected in such a search can be determined from the distribution for
the power expected at a single frequency. We work out the form of this
distribution in the next section.

\section{Extreme value statistics}
\label{sec:stat}

Our challenge in evaluating the evidence for periodic emission from
GRBs and SGRs is to interpret the values of the Rayleigh power obtained
from each non-repeatable burst event. Since we have designed our
frequency search to avoid contamination by the slowly varying burst
envelope, the (marginal) probability distribution for the Rayleigh
power should be the same for each frequency in our search if there is
no rapid oscillation of the source.  Thus, for a particular burst
record the most promising frequency candidate is the one at which the
largest value of the power is found, $R_{\rm max}$. This maximum must
be compared with the distribution of maxima expected under the null
hypothesis of an uniform phase distribution.

In general, if we have $N$ samples $\{x_i\}$, $i=1$ to $N$, 
drawn independently from a probability density function $f(x)$ 
with a cumulative distribution function
\begin{equation}
F(x) = \int^{x} \! ds\ f(s),
\label{}
\end{equation}
and we have determined the largest of the $x_i$, $X$, then
the probability distribution $g(X)$ 
for the extreme value $X$ is (Ochi 1990)
\begin{equation}
g(X) = N f(X) F(X)^{N-1}.
\label{}                                     
\end{equation}
The
most probable value of $X$
can be found by solving
\begin{equation}
\frac{f'(X)}{f(X)}= -(N-1)\frac{f(X)}{F(X)}.
\label{cond}
\end{equation}
For the Rayleigh power, $R$, the cumulative distribution at a single
frequency under the null hypothesis is $F(R) = 1 - e^{-R}$, and
the density function is $f(R) = F'(R) = e^{-R}$.  If $R_{\rm max}$ is
the maximum among $N$ independent samples of $R$, its density is
\begin{equation}
g(R_{\rm max}) = N e^{-R_{\rm max}} (1-e^{-R_{\rm max}})^{N-1}.
\label{maxdist}
\end{equation}
Using \ceq(cond)
the most probable value of $R_{\rm max}$ is
\begin{equation}
\hat{R}_{\rm max} = \ln N ;
\label{eq8}
\end{equation}
and
\begin{equation}
\everymath{\displaystyle}
  \begin{array}{rl}
        G(R_{\rm max})  &= \int_0^{R_{\rm max}} \! \! dr\ g(r)  \\
			\rule{0in}{5ex}
                        &= (1-e^{-R_{\rm max}})^{N} \; 
  \end{array}
\end{equation}
is the cumulative distribution for $R_{\rm max}$. For
$Ne^{-R_{\rm max}} \ll 1$, $1-G(R_{\rm max}) 
\approx Ne^{-R_{\rm max}}$.

The above analysis assumes the $N$ samples are independent.  When one
is oversampling, this is not the case.  But in such situations a good
approximation to the extreme value distribution can often be found by
using for $N$ some number below the actual number of (dependent)
samples; this is the {\em effective} number of samples.  In our case,
we sample 13194 frequencies for every record regardless of the
duration, $T$, and number of events, $n_{\rm dat}$ in the record.  We
thus expect the effective number of frequencies sampled, $N_{\rm eff}$,
to vary from trigger to trigger as a function of $T$ and $n_{\rm
dat}$.

In order to find a function which accurately describes this 
relationship a number of burst events were simulated varying $T$ and
$n_{\rm dat}$, and $N_{\rm eff}$ was calculated by a maximum likelihood 
fit to simulation results for the number of (effectively)
independent samples, $N$. The method used to simulate data is 
discussed in the next Section. 
For each simulated burst, $i$, we determine a maximum Rayleigh 
power $R_{m,i}$. 
The likelihood 
for the effective number, $N$, given the set $\{R_{m,i}\}$
is
\begin{eqnarray}        
     \mathcal{L} (\mbox{$N$}) & = & \prod_{i}^{N_b} g(R_{m,i}, N) \nonumber \\
                     & = & \prod_{i}^{N_b} Ne^{-R_{m,i}}(1-e^{-R_{m,i}})^{N-1},
\end{eqnarray}
where $N_{b}$ is the number of bursts or simulated events
in our sample. We can
determine the value of $N$ which  maximizes $\mathcal{L}$($N$); this 
is 
\begin{equation}
N_{\rm eff}=-\frac{N_{\rm b}}{\ln[\prod_{i}^{N_{\rm b}} (1-e^{-R_{m,i}})]},
\label{Neff}
\end{equation}
the effective number of frequencies we use when 
calculating the significance of the maximum Rayleigh power.                    

We simulated sets of 100 burst events  
for 29 combinations 
of $T$ and $n_{\rm dat}$, and calculated $N_{\rm eff}$ 
for each. 
Figure 2 shows $N_{\rm eff}(n_{\rm dat})$ 
for $T=1.1$ and $T=2.0$ s. We can see 
that $N_{\rm eff}$ does not depend on $n_{\rm dat}$ 
systematically. Figure 3 shows the values of 
$N_{\rm eff}$ as a function of $T$ for our simulations.
The smooth curve is the best fit obtained 
for a functional form
\begin{equation}
N_{\rm eff}(T)= N(1-e^{-aT-bT^{2}}),
\label{fit}
\end{equation}
where $N$ is the actual number of frequencies sampled and $a$ 
and $b$ are fit parameters found to be $a=0.543$ and $b=0.202$. 
The form of this equation was chosen to 
satisfy $N_{\rm eff} = 0$ at $T=0$ and $N_{\rm eff} \rightarrow N$ 
as $T \rightarrow \infty$. 

\begin{figure}
\plotone{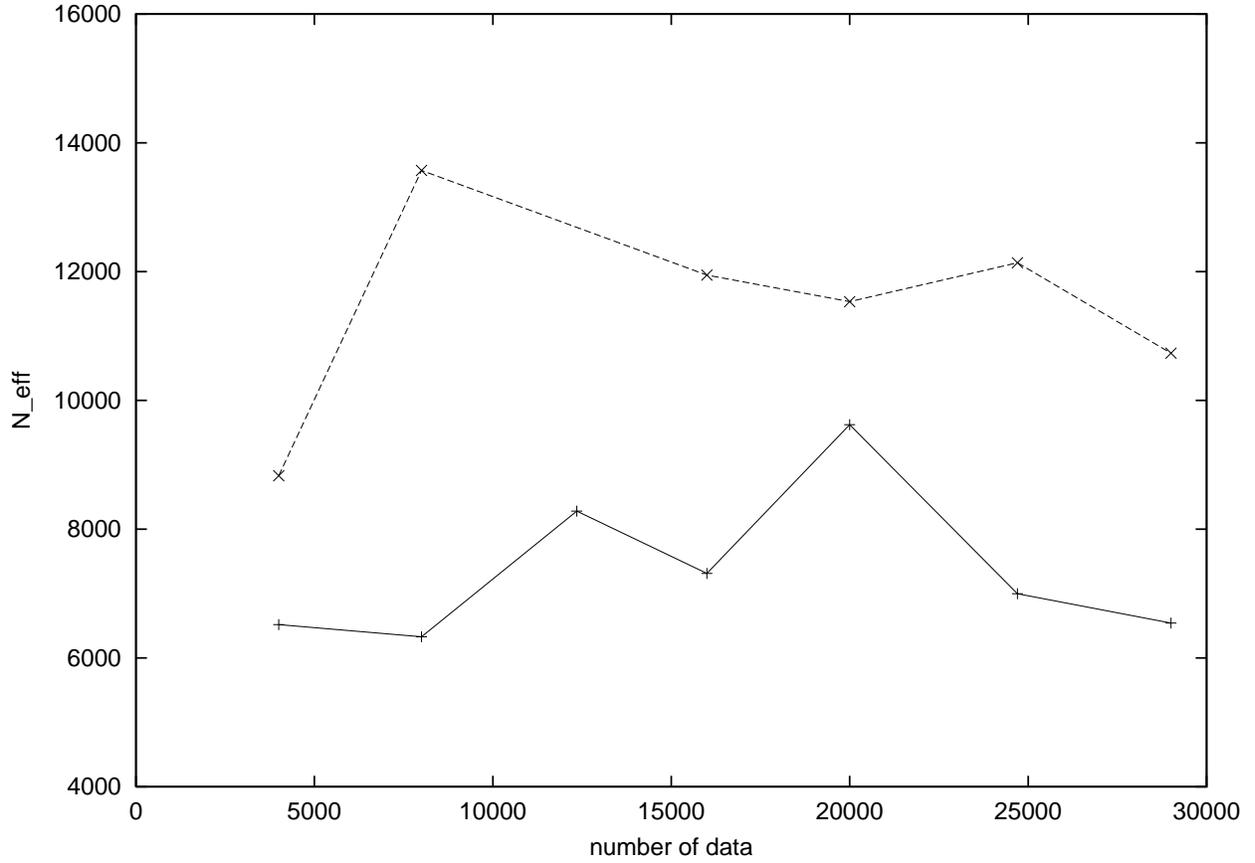}
\caption{Calculated value of the 
effective number of frequencies sampled, $N_{\rm eff}$, 
vs.\ the number of arrival times comprising 
the data record, $n_{\rm dat}$,
obtained from simulated data using a uniform 
rate function with duration, $T=1.1$ s (bottom) and 
$T=2.0$ s (top) and using 13194 frequencies from
400~Hz to 2500~Hz. There is no appreciable
dependence of $N_{\rm eff}$
on the number of data $n_{\rm dat}$.}
\end{figure}

\begin{figure}
\plotone{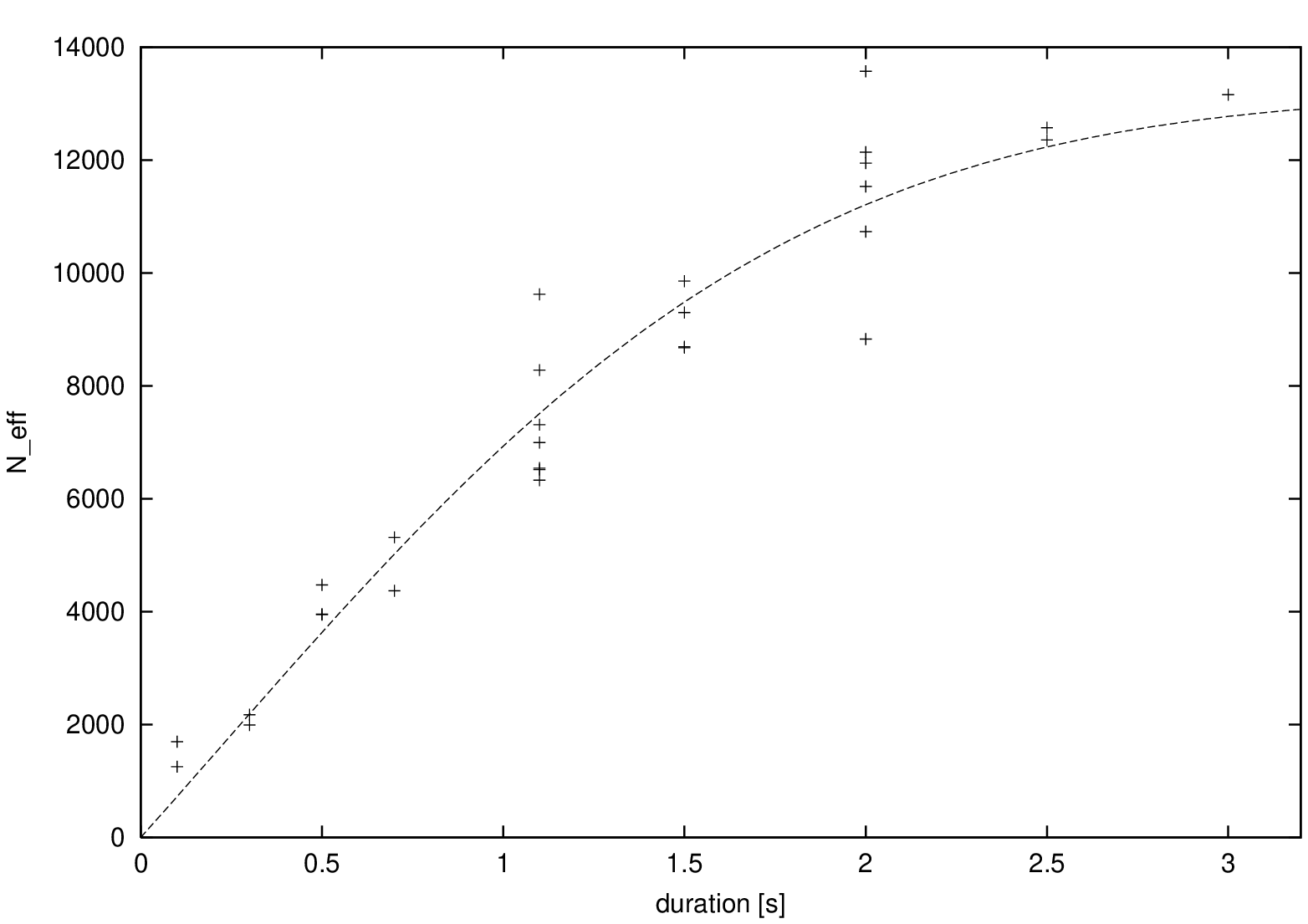}
\caption{Effective number of frequencies sampled, $N_{\rm eff}$,
vs.\ record duration, $T$. Points  
obtained from simulated data using a uniform rate function
and records with durations between 0.1~s and 3.0~s,
and examining 13194 frequencies from
400~Hz to 2500~Hz. Plotted 
with best-fit function obtain from least squares fitting
and 2 parameter model of the form, $N_{\rm eff}(T) =
N(1-e^{-aT-bT^{2}})$, where $a=0.543$ and $b=0.202$ are
the fit parameters.}
\end{figure}

\section{Simulations and detection}
\label{sec:simdet}

In this section we discuss the method used to 
simulate burst data and analyze simulated arrival times 
to better understand our ability to detect periodicities 
using the Rayleigh test. A commonly used model for burst
light curves is the fast-rise, exponential 
decay (FRED)  model. We adopt this basic form here and model
a burst as a 
two-sided exponential function plus 
background, with some percentage of the flare or 
background modulated at a frequency $f$. The intensity 
function we use for simulations is

\begin{equation}
\everymath{\displaystyle}
	I(t) = \left\{  \begin{array}{ll}
		Ae^{\Gamma_{1}(t-t_{p})}[1+\varepsilon_{1} \cos(2 \pi f(t-t_{p}))]
		+ B[1 + \varepsilon_{2} \cos(2 \pi f (t-t_{p}))]
		& t \leq t_{p} \\
		\rule{0in}{5ex}
		Ae^{-\Gamma_{2}(t-t_{p}}[1+\varepsilon_{1} \cos(2 \pi f(t-t_{p}))]
		+ B[1 + \varepsilon_{2} \cos(2 \pi f (t-t_{p}))] 
		& t  >t_{p}.
			\end{array} 
	       \right.
\label{intensity}
\end{equation}	
The input parameters are the record duration, $T$, 
time of peak intensity, $t_{p}$, flare amplitude, $A$, 
rise constant, $\Gamma_{1}$, decay constant, $\Gamma_{2}$, 
background amplitude $B$, pulsed fraction on flare, 
$\varepsilon_{1}$, pulsed fraction on background, 
$\varepsilon_{2}$, and the frequency of the periodic 
emission component $f$. 
In principle, we could have different frequencies and
phases for signal and background; in simulations, we
generally took either $\varepsilon_{1} = 0$ or
$\varepsilon_{2} = 0$.
For simplicity we choose the 
phase to be zero at $t=t_{p}$. This function is continuous 
across $t_{p}$ where it reaches a maximum,
\begin{equation}
I(t_{p})=A(1+\varepsilon_{1})+B(1+\varepsilon_{2}).
\end{equation}
The simulations that were done to determine the value of 
$N_{\rm eff}(T)$ used data generated from a constant 
intensity function ($A=\varepsilon_{2} = 0$). This was 
done after first verifying that the existence of a FRED-type 
flare did not affect the value of $N_{\rm eff}$ that 
we obtain. To establish this we generated 500 simulated burst 
records with roughly 29000 points each; assuming
$T=1.5, t_{p}=0.4, A=15, 
\Gamma_{1}=25.0, \Gamma_{2}=20.0$, and $B=12$, and with no 
periodic component. 
The value obtained for $N_{\rm eff}$ 
using equation (\ref{Neff}) from these bursts was $N_{\rm eff} 
= 9856$. The calculation for simulated data arising from 
a constant intensity for the same duration and approximately 
the same number of accepted times led to $N_{\rm eff} = 9299$. 
A similar calculation was done for $T=0.5$ and the difference 
between the obtained $N_{\rm eff}$ values 
was less than 1\% with the data derived from the constant 
intensity function leading to a larger value.
These differences in $N_{\rm eff}$ are small
and do not significantly change the formal significance levels one 
would ascribe to an observed $R_{\rm max}$, justifying our
use of the simplest null hypothesis for the calculations
of section 6.  This is not surprising given our previous
finding that for frequencies above 400~Hz, typical burst
envelopes do not change the expected power distribution
from that found using a uniform phase distribution.

In addition to using simulations to establish expected 
distributions of $R_{\rm max}$ values we can also use 
simulated data to establish 
how strong the
periodic component in our model must be to be detectable.
The significance of 
a particular maximum of the Rayleigh power is given by 
equation (\ref{significance1}). We can use
this expression and simulated data with a known 
input frequency and known pulsed fraction
to determine what percentage of these records 
lead to maximum Rayleigh powers at a particular significance level. 
We did not anticipate that the results would depend on the 
value of the input frequency chosen, but to test this we have 
conducted simulations with two different input frequencies:
800 Hz and 2000 Hz. For each of these frequencies we generated 
100 simulated data sets using equation (\ref{intensity}) 
with 
$T=2.0$, $t_{p}= 0.4$, $A=20$, $\Gamma_{1} = 10$, $\Gamma_{2} = 10.0$,
$B = 5.0$, $\varepsilon_{2} = 0$, and 
$\varepsilon_{1} = 0.07$, $\varepsilon_{1} = 0.17$, 
and $\varepsilon_{1} = 0.35$
and then searched the simulated data for periodicity.
The corresponding pulsed fractions are 0.04, 0.10, and 0.20 respectively. 
In Figure 4 we show scatterplots of Rayleigh power
vs.\ $\varepsilon_{1}$ for each set of 100 simulations.  Figure~4{\em
(a)} shows the results for an 800 Hz input frequency and Figure~4{\em
(b)} shows results for 2000 Hz.  Clearly when $\varepsilon_{1}$ (and
hence the pulsed fraction on the flare) is increased the likelihood of
a significant detection increases. With the input parameters chosen
here and $\varepsilon_{1}=0.35$ (pulsed fraction of 20\%), we find
Rayleigh power values that would be considered statistically
significant in all of the 200 simulations. A value of
$\varepsilon_{1}=0.17$ (pulsed fraction of 10\%) leads to significant
detection in more than 50\% of the trials.  These simulations indicate
that if the emission from a typical burst has an oscillatory component
at the 10\%-20\% level, then the method that we employ here would allow
us to confidently reject the hypothesis of a uniform distribution of
phases and claim a significant signal detection (at the correct
frequency) in a majority of cases.  This conclusion does not appear to
depend on the frequency of the oscillations, provided it is large
enough.

\begin{figure}
\epsscale{0.8}
\plotone{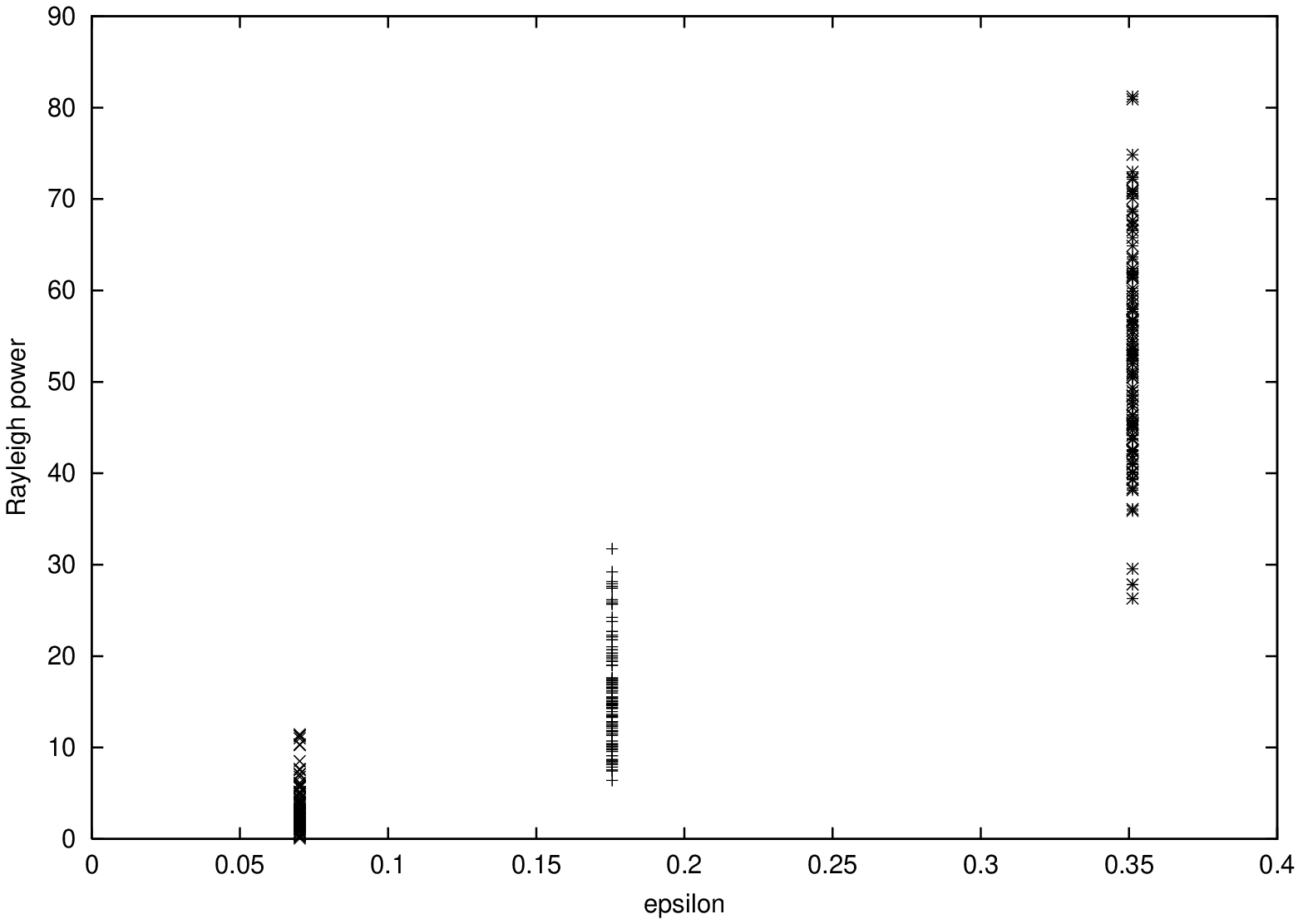}
\plotone{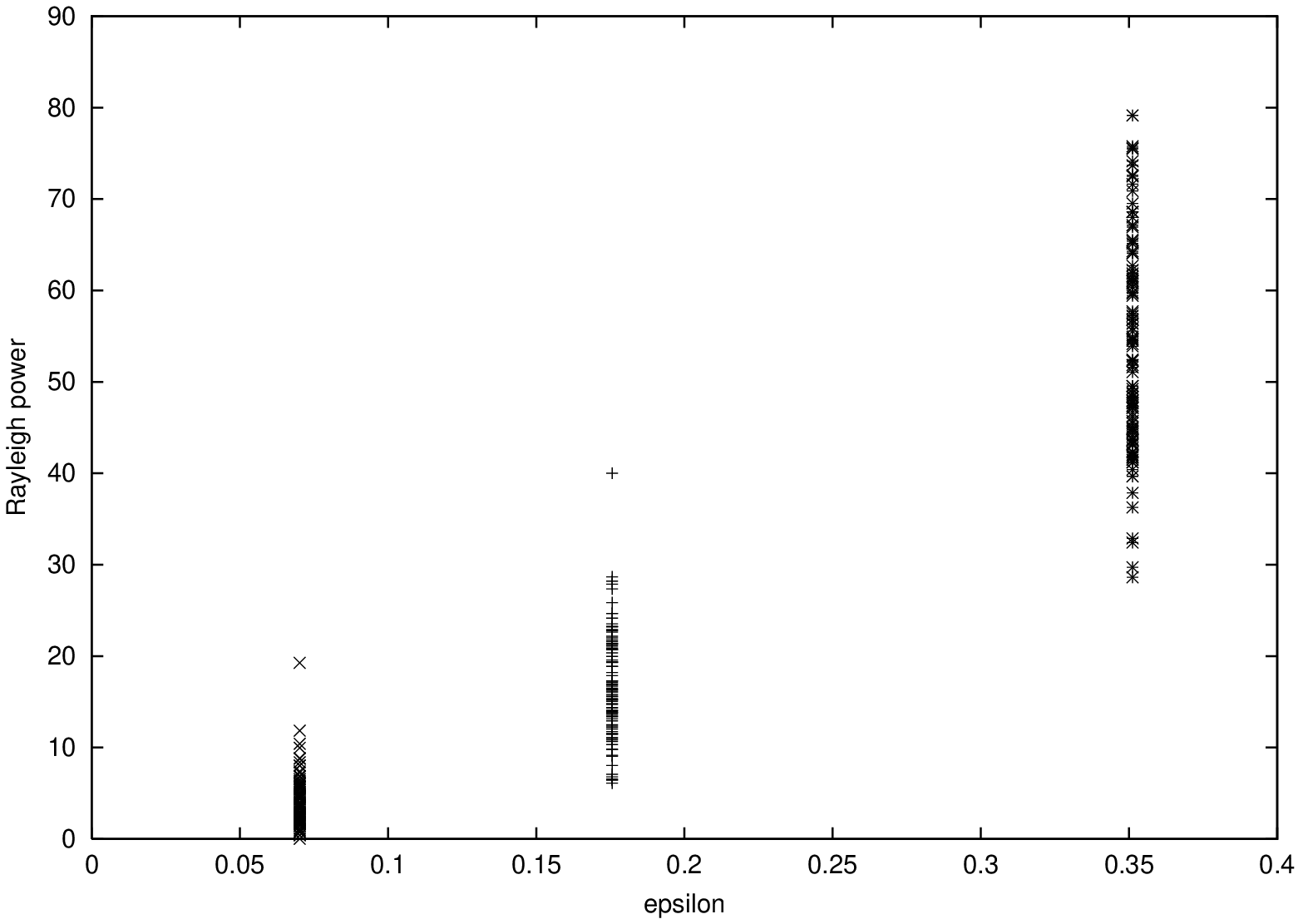}
\epsscale{1.}
\caption{The Rayleigh power calculated at the input frequency in
simulations vs.\ $\varepsilon_{1}$ of equation (\ref{intensity}).  The
upper panel shows the power calculated for 800~Hz, and the lower panel
shows the power calculated at 2000~Hz.  These simulations demonstrate
that with $\varepsilon_{1}=0.17$ (total pulsed fraction of 10\%) we
detect a signal with significance, $s \leq$1\% in more than half the
trials, and for $\varepsilon_{1}=0.34$ (total pulsed fraction of 20\%)
we have significant detection at the correct frequency in 100\% of the
simulations.}
\end{figure}

\section{Analysis}
\label{sec:analysis}

In this section we present the results of our analysis 
of the BATSE TTE data. In order to evaluate the evidence for 
periodic emission in the 2203 GRB and 152 SGR flare records 
in our study we have calculated the Rayleigh power for 13194
frequencies from 400 Hz to 2500 Hz for each trigger. 
Because we consider all frequencies equally likely a priori, 
the best candidate frequency is the 
one at which the maximum power is attained. 
In order to judge the significance of a particular $R_{\rm max}$ 
value obtained from a burst record, we use the $R_{\rm max}$ 
distribution in equation (\ref{maxdist}). The tail area of 
this distribution is the probability of obtaining a larger 
value without the presence of a signal. The number of frequencies 
that enters this equation is given by equation (\ref{fit}) as a 
function of the time coverage $T$. 
Before doing our analysis we decided that we 
would keep the two populations of bursts, SGRs and GRBs, separate 
as they represent different 
classes of physical phenomena. We assign a score to
each $R_{\rm max}$ value, given by
\begin{equation}
s  = 1-(1-t)^{N_{\rm b}} ,
\label{significance1} 
\end{equation}
where $t=[1-(1-e^{-R_{\rm max}})^{N_{\rm eff}}]$, $N_{\rm b}$ is 152
for SGRs and 2203 for GRBs.  This score is the significance that would
be associated with $R_{\rm max}$ were it the largest value in the
population.  This interpretation applies only to the single largest
$R_{\rm max}$ in each population, but $s$ is useful more generally to
identify potentially interesting events (those with small scores).  Of
the 2203 classical bursts in our study and 152 flares from SGR sources,
only 4 led to Rayleigh powers high enough to have scores $s \leq 0.6$.
Two of the records are GRB triggers, these are trigger 1493 with
$R_{\rm max} = 15.85$ occurring at 819.0 Hz with $s=0.54$  and trigger
2101 with $R_{\rm max} = 18.12$ found at 878.7 Hz and $s=0.32$.  The
other two are SGR records: trigger 6855 with $R_{\rm max} = 16.78$
found at 872.0 Hz with $s=0.10$, and trigger 7041 with $R_{\rm max} =
18.26$ occurring at 2496.7 Hz with $s=0.01$.  If one considers these
results purely from the standpoint of statistical significance, it
would seem that only trigger 7041 warrants additional attention, but
more can be said about triggers 2101 and 6855.

The TTE light curve for trigger 2101 is shown in the inset of Figure~5.  
The most conspicuous feature of this trigger is a rapid drop in the
number of counts recorded at around 0.7 seconds.  No such feature is
present in light curves produced using other (binned) BATSE data types
for this event, indicating a problem with the TTE data for this burst.
The TTE record begins with roughly 11000 photons that arrive at a rate
consistent with background as observed in other records, roughly 20 per
ms; but the rate falls abruptly to the anomalously low level of just
one per ms. Because of the low count rate, TTE data exist out to 7~s.
Because this record produced a very large Rayleigh score and the light
curve revealed the odd change in count rate, we divided the record into
two sections and calculated the Rayleigh power for each in an attempt
to learn if the power arose from the first part of the record or the
later low-rate portion.  We find that the origin of the large Rayleigh
power is the low-rate portion of the record. If we take all of the
counts with $t>0.75$~s and calculate the Rayleigh power spectrum we
find the maximum to be at 902 Hz, with a Rayleigh power of 29.7 shown
in the Figure, an extremely significant value. In addition to the peak
at 902 Hz, there are other peaks with very large powers at other nearby
frequencies, separated by roughly 10~Hz. It is unclear what causes
these large values, but the anomalous feature of the TTE light curve
indicates a serious problem with these data.

\begin{figure}
\plotone{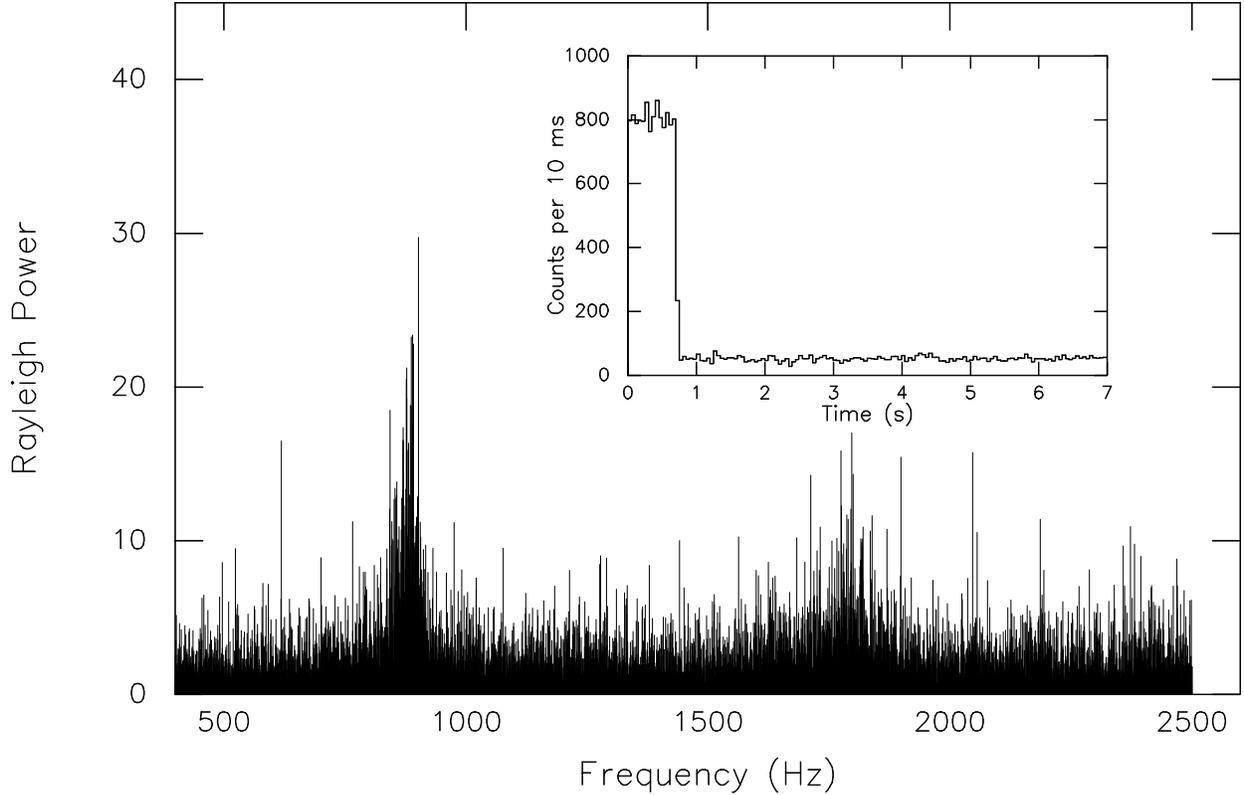}
\caption{Rayleigh power as a function of frequency calculated
for the low count rate($t> 0.75$~s) portion of trigger 2101. Note
the large power observed near 900 Hz and 1800 Hz. 
Similar power spectra were observed from other triggers where
anomolously low count rates were recorded. The light curve generated
from TTE data for this event is shown in the inset. Since this
light curve is inconsistent with that produced from other archived
data, we conclude that this set of TTE data is corrupted and that
the large power here is not evidence of a periodic signal.}
\end{figure}

A distinguishing characteristic of this record is its long time
coverage.  In order to further study this phenomenon we searched the
data sets for other records with atypical durations. We found among the GRB
records four which have anomalously long coverage. The four GRB records
occur in succession; they are triggers 2101, 2103, 2104 and 2105.
Trigger 2102 had already been rejected from our study due to a problem
with the data; the end of the record consists of many events all
assigned the same arrival time. Inspection of the TTE light curves of
the four GRB profiles indicates that each of them has the same basic
appearance---a fast drop in count rate at about 1~s and in each case
the TTE light curve bears no resemblance to light curves based on other
BATSE data types.  For each of these records we have separated out the
low-count rate portion of the record and calculated the Rayleigh power
at the frequencies in our search range.  The results demonstrate that
the source of the significant power in trigger 2101 is also present in
the other records.  We cannot say what the cause of this signal is, but
we are confident that it is not astrophysical, and that the significant
power that we find in these records does not constitute evidence of
periodic emission from these GRBs.

One SGR record also has anomalously large time coverage of 6.5~s. 
This trigger is 6855, the burst with data leading to the second 
most significant maximum Rayleigh power among SGRs. 
Although the light curve obtained from this record did not 
reveal the rapid decrease in count rate found in the records 
mentioned above, we note that the record indicates a low count rate 
during the entire observation time. The frequency at which the 
power is a maximum is 872.0~Hz, similar to what we 
found in the 
problematic GRB triggers, 2101, 2103, 2104, and 2105. 

The burst trigger from which we obtained the most significant evidence
for periodicity in our study is trigger 7041. The light curve for this
SGR outburst is shown in the inset in Figure~6.  The burst is fully
captured by the TTE data and there are no evident data anomalies in
this record. The record for this burst contains 22383 arrival times
and the time coverage is 1.0 s. The frequency which corresponds to the
maximum Rayleigh power for this burst is 2496.7~Hz. Its Rayleigh power
of 18.3 corresponds to a score of $s=0.0124$; since this burst has the
largest power, this score is the significance level. The source of this
flare is SGR~1900$+$14, the same source that just 4 days before trigger
7041 produced a huge burst which was observed to have a 5 s periodicity
(Hurley et al. 1998). This large flare was not captured by BATSE as the
source was unfortunately occulted by the Earth at the time.  There
were 5 other triggers from SGR~1900$+$14 before 7041 but after the large
flare of 27 Aug 1998, and 7 that followed in the subsequent 2 days.
Because the maximum was obtained so near to our high-frequency search
limit, and so near the large flare of 27 Aug, we were concerned that
we might be missing large Rayleigh values in these other records by ending
our search before finding them. To check this we extended our search to
higher frequencies (up to 5000~Hz) for the triggers from SGR~1900$+$14 that
occur after the large flare. In these additional calculations we found
no evidence for substantial power near 2497~Hz. Although the significance
level associated with possible emission from trigger 7041 at 2496.7~Hz
is interestingly small, and the folded light curve has a realistic
appearance (Figure~7), we feel that this evidence is not strong enough
to justify a claim of discovery of periodic emission.  Other flares
from the same source that occurred just before and soon after this
trigger do not exhibit the same periodicity.  We consider this record
as an interesting oddity---a statistical anomaly rather than a
bona-fide periodic signal.

\begin{figure}
\plotone{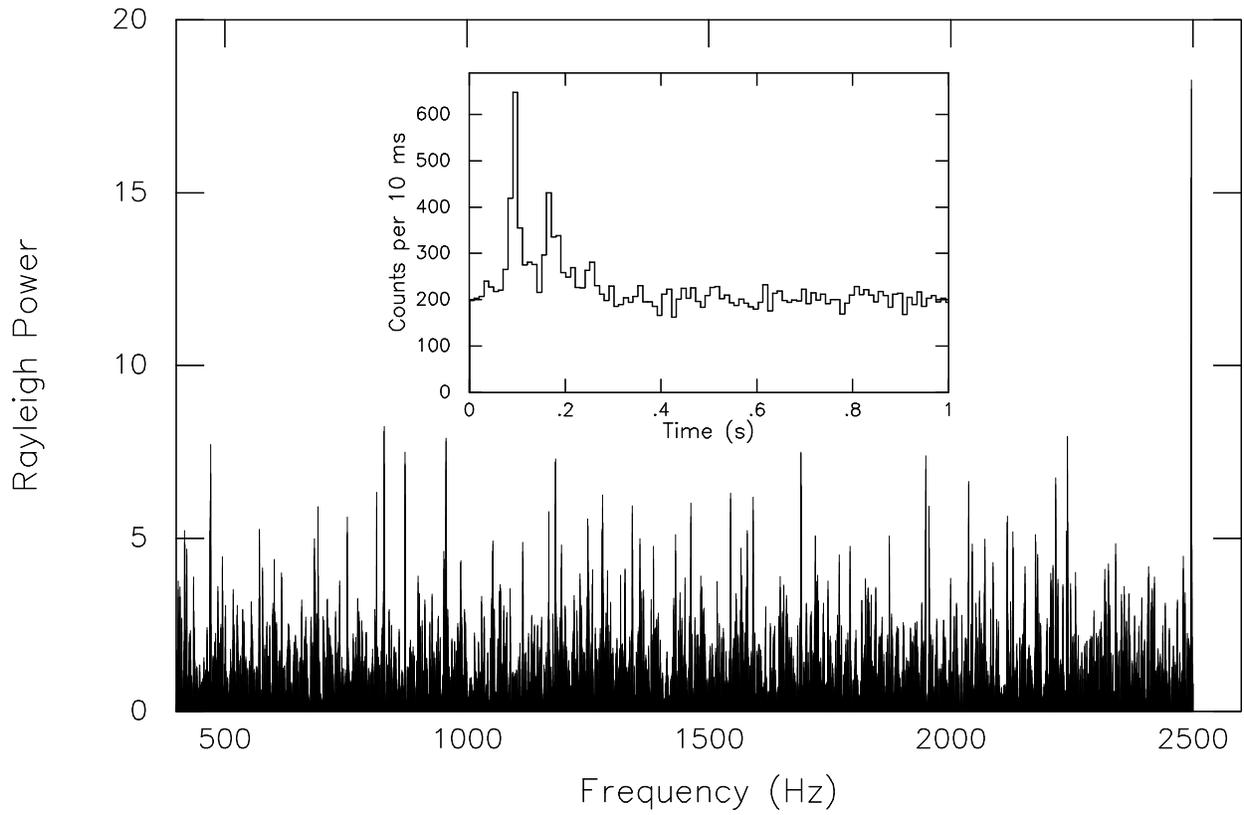}
\caption{Rayleigh power spectrum for trigger 7041.  A Rayleigh power of
18.3 is calculated for this trigger at 2497~Hz. This value results in a
detection significance of 1.25\% and is our most significant candidate
periodicity.}
\end{figure}

\begin{figure}
\plotone{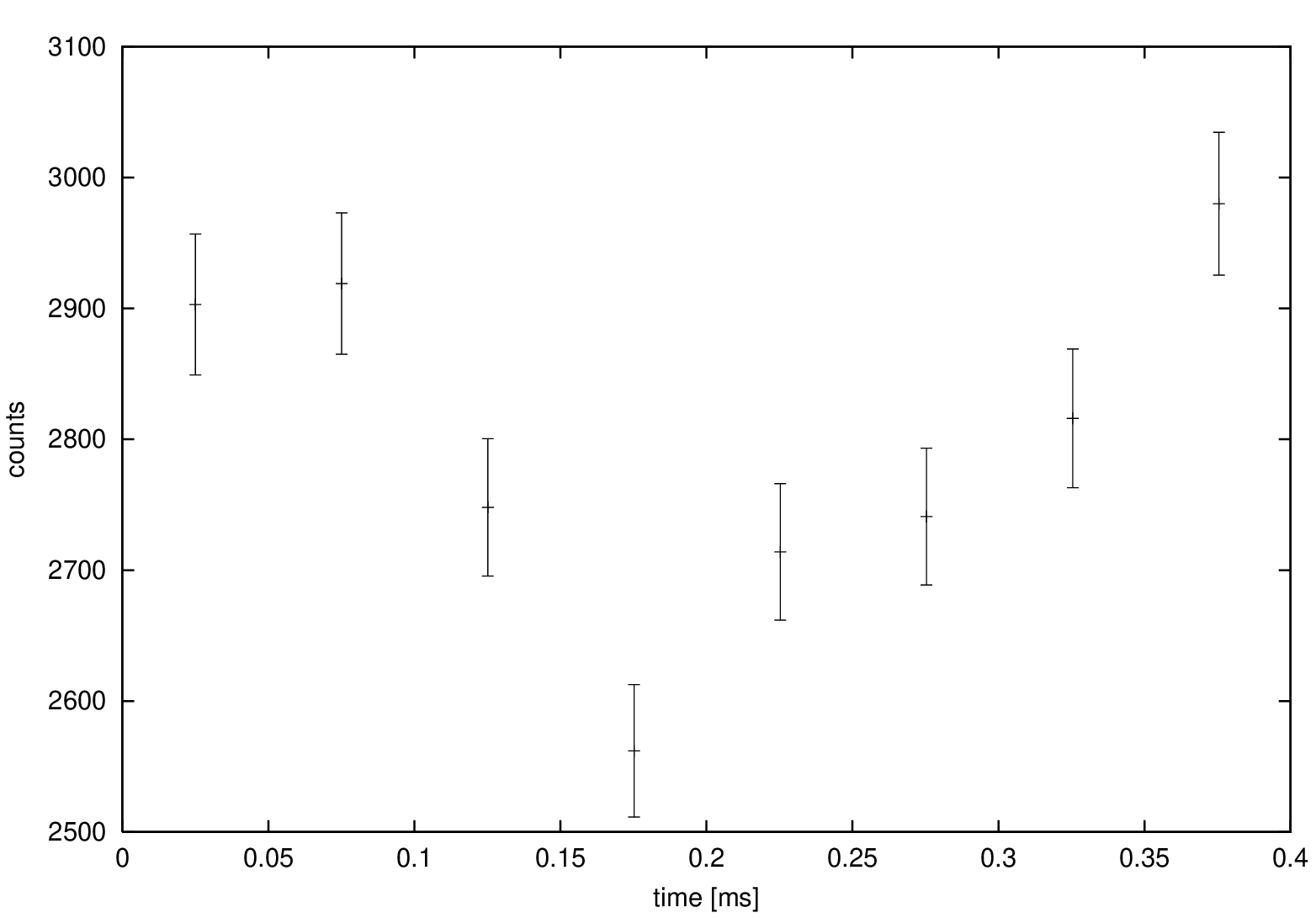}
\caption{Folded light curve for data from trigger 7041 for frequency
$f=2497$~Hz ($0.4$~ms period). If there is a periodic component in this
record, then the total pulsed fraction of the signal comprises roughly
10\% of the data.}
\end{figure}

\section{Conclusions}
\label{sec:conc}

This study is the first extensive search for high-frequency 
periodic emission to use the BATSE Time-Tagged Event data. 
The outcome of this study has not been the discovery of 
high-frequency periodicities in the burst records from SGRs 
and GRBs, but instead absence of evidence of these signals.
In our study we have found just one plausible candidate for 
periodic emission at a frequency in our search range. 
The Rayleigh power obtained from trigger 7041 corresponds to 
detection near the 1\% significance level, but the lack of 
emission at the same frequency in bursts from the same source 
(SGR~1900$+$14) at nearly the same time argues against this as a true 
detection. We have uncovered evidence for occasional
TTE data corruption that produces a signal with $ f \approx 900$~Hz;
but the corrupt data is identifiable by obvious discrepancy 
between the TTE light curves and light curves using 
other data types.

We have simulated burst data in an effort to determine how strong a
periodic component must be in the signal from a burst to be detectable.
A rough rule-of-thumb is that for a catalogue with $\approx 2000$
bursts, periodic modulation comprising $\approx 10$\% of the photons in
a given BATSE TTE record would have yielded significant Rayleigh power
about half the time. (Since background is included in the record, the
pulsed fraction in the burst signal must be higher.) Our analysis found
oscillations at this level in only one TTE record, trigger 7041 from
SGR 1900+14. That so few significant values of the Rayleigh power were
found indicates that there is no substantial high-frequency periodic
emission from these sources.

\acknowledgments

We gratefully acknowledge helpful conversations and correspondence
about BATSE TTE data with Jay Norris, Jeff Scargle, and Jerry Bonnell.
This work was supported by NASA grants NAG~5-3800, NAG~5-8356, and 
NAG~5-3427.  This paper forms part of A.~Kruger's M.S.\ thesis
at Cornell University.

\clearpage

\end{document}